\documentclass[preprint,showpacs,preprintnumbers,amsmath,amssymb]{revtex4}
\usepackage{graphicx}% Include figure files
\usepackage{dcolumn}% Align table columns on decimal point
\usepackage{bm}% bold math

\begin{document}

%\preprint{APS/123-QED}

\title{Coherent Dissociation $^{12}$C~$\rightarrow$~3$\alpha$ in Lead-Enriched Emulsion at 4.5 GeV/$c$~per Nucleon}

\author{V.~V.~Belaga}
   \affiliation{Joint Insitute for Nuclear Research, Dubna, Russia}  
\author{A.~A.~Benjaza}
   \affiliation{Aden University, Aden, Republic of Yemen}  
\author{V.~V.~Rusakova}
   \affiliation{Joint Insitute for Nuclear Research, Dubna, Russia}
\author{J.~A.~Salamov}
   \affiliation{Insitute of Nuclear Physics, Tashkent, Uzbekistan}
\author{G.~M.~Chernov}
   \affiliation{Joint Insitute for Nuclear Research, Dubna, Russia}
   
\date{\today}% It is always \today, today,

\begin{abstract}
\indent The transverse-momentum distributions and correlation characteristics of relativistic $\alpha$~particles from the coherent dissociation of a carbon nucleus into three $\alpha$ particles at 4.5 GeV/$c$~are studied in lead-enriched emulsion. Comparative analysis of data obtained in ordinary and lead-enriched emulsion stacks is performed. It is shown that the statistical model of rapid fragmentation does not describe the momentum and correlation characteristics of a $\alpha$ particles in the rest frame of the carbon nucleus. The estimated decay temperature of $^{12}$C is weakly dependent on the target atomic mass. It is shown that the carbon nucleus undergoing fragmentation acquires angular momentum in the collision.\par
\indent \par
%\indent DOI: 10.1134/S1063778810120161\par
\end{abstract}
  %    {PACS-key}{21.45.+v} \and
   %   {PACS-key}{23.60+e} \and
    %  {PACS-key}{25.10.+s}  

\maketitle

\section{\label{sec:level1}}

\indent As far back as 1953, Pomeranchuk and Feinberg \cite{Pomeranchuk} predicted the existence of a special type of collisions involving complex nucley. Collisions of this type do not lead to the disintegration or excitation of a nucleus that causes the dissociation of a high-energy incident hadron and are characterized  be low momentum transfer and zero charge exchange. These extremely peripheral reactions can proceed through Pomeron exchange (diffraction mechanism) or photon exchange (Coulomb mechanism). The second mechanism is dominant at large nuclear charges. Coherent production of particles attracts the attention of researches because of the relative simplicity of theoretical description and because of some interesting features peculiar to such processes $-$~for example, the remarkable property of a nucleon ensemble to enhance or to suppress inelastic diffraction mechanisms, or the unique possibility to study the interaction of quasi-stable hadron resonances and clusters with intra-nuclear nucleons.\par
\indent Such reactions can occur in nucleus-nucleus interactions, in which case a projectile is also a nucleus. Here, coherent interaction leads to nuclear fragmentation rather than to the production of new particles. Reactions of this type usually have comparatively high energy thresholds and, hence, belong to the realms of high- energy nuclear physics. Indeed, applying the energy-momentum conservation law to the dissociation of a projectile nucleus $A$ with mass $M_{0}$ and momentum $p_{0}$ into the system of $n$ fragments with masses $m_{i}$($i$~=~1,..., $n$) in a collision with a target nucleus $B$, we have \cite{Chernov} \par

\indent $$\frac{M^{*2}-M^2_0}{2p_0}+\frac{E_0+M}{2Mp_0}q^2,$$\\ where $M^*$ is the effective mass of the system of $n$ fragments, and $q$ is the 3-momentum transfer in the collision. The coherence condition 1/$q~>~R$~($R$ is the radius of the target nucleus) leads to the following constraint on the longitudinal component of the momentum transfer in inelastic momentum diffraction \cite{Pomeranchuk}: \par

\indent $$q_L<\mu/B^{1/3}.$$\\ Here, $B$ is the mass number of the target nucleus, and $\alpha$ is the pion mass. Considering that the effective mass has a minimum value of $M^*_{min}~=~\sum_{i=1}^nm_i$ and using the above formulas, we find that the threshold of the coherent dissociation of the nucleus is given by\par
\indent $$p^{min}_0\cong\frac{(\sum_{i=1}^nm_i)^2-M^2_0}{2\mu}B^{1/3}\cong\frac{M_0B^{1/3}}{\mu}\Delta,$$ \\ 
where $\Delta~=~\sum_{i=1}^nm_i~-~M_0$ is the mass defect with respect to the dissociation channel under consideration. The coherent dissociation of nuclei is an important source of information about the internal structure of nuclei undergoing fragmentation, because their decays can be studied only at very low energy-momentum transfer and without a threshold of fragmentation product detection. In this connection, it is surprising that investigation of inelastic coherent dissociation of relativistic projectile nuclei only begins to gain momentum.\par
\indent This article reports the results obtained by studying the dissociation of $^{12}$C nuclei with initial momentum 4.5 GeV/$c$ per nucleon in interactions with lead-enriched photoemulsion. Earlier \cite{bvdklmt,Abdurazakova}, searches for the coherent reaction $^{12}$C~$\rightarrow$~3$\alpha$ at the same initial momentum were performed in ordinary emulsion. Apart from these emulsion studies, we are aware of only two investigations performed at relativistic energies \cite{Engelage,Bondarenko}. Engelage $et~al$. \cite{Engelage} used a HISS spectrometer at the accelerator Bevalac and a carbon target to study $^{12}$C dissociation. Bondarenko $et~al$. \cite{Bondarenko} attempted to select coherent decay into three a particles from inelastic interactions of carbon nuclei at $p_0~=~4.2$~GeV/$c$~per nucleon in the 2-meter propane bubble chamber of the Laboratory of High-Energy Physics at JINR.\par
\indent The search for reactions with three relativistic charged fragments of the projectile with charge $z$~=~2 was performed by scanning along tracks and over areas in the stack of nuclear emulsions enriched in lead to a concentration of 1 Pb atom per 5 Ag atoms. Salts of lead were introduced into standard liquid BR-2 emulsion immediately prior to the preparation of photo-emulsion layers. The crystal sizes of the salts in the prepared emulsion layers were commensurate with the sizes of Ag Br crystals (0.3 $-$ 0.5 $\mu$m). The layer thickness was about 500 $\mu$m. Comparative nuclear compositions of the standard BR-2 emulsion and that used in our experiment are given in Table 1 \cite{Akhrorov}.\par

\begin{table}
\caption{\label{}}
%\lable{Table:1}
\begin{tabular}{c|c|c|c|c|c|c|c}
\hline\noalign{\smallskip}
& \multicolumn{7}{r}{Number of nuclei in 1 cm$^3$ ($\times10^{22}$)}  \\
\noalign{\smallskip}\hline\noalign{\smallskip} 
Photoemulsion & H & C & N & O & Br & Ag & Pb \\
\hline\noalign{\smallskip}
BR-2 (standart emulsion) & 2.97 & 1.40 & 0.37 & 1.08 & 1.03 & 1.03 & $-$ \\
\noalign{\smallskip}\hline\noalign{\smallskip} 
BR-2+Pb (this study) & 3.26 & 1.64 & 0.28 & 1.49 & 0.76 & 0.76 & 0.15 \\
\hline\noalign{\smallskip}
\end{tabular}
\end{table}

\indent The stack was irradiated by a $^{12}$C ion beam at a primary momentum of 4.5 GeV/$c$~per nucleon from the synchrotron of the Laboratory of High-Energy Physics at JINR. Scanning along tracks was carried out to evaluate the mean range (i.e., cross section) for the reaction under investigation in the emulsion stack and for the subsequent comparison of this range with the corresponding quantity for ordinary emulsion \cite{Abdurazakova}. Scanning over areas, which leads to much faster acquisition of events we are interested in, was carried out to increase the statistics of these rare events.\par
\indent Events that involve three well identified doubly charged relativistic fragments of a projectile nucleus and which were selected under suitable geometric conditions were measured by a special method that rules out the effect of distortions on measurements of angles. As in \cite{bvdklmt,Abdurazakova} we disregarded the possible admixture of $^3$He nuclei among fragments with $z$~=~2 and considered them as $\alpha$ particles. Events selected for measurements must have no indication of target excitation or breakup (\lq\lq pure stars\rq\rq) and addition relativistic tracks. In all, we found and measured 72 such events. We estimated the mean range with respect to $^{12}$C dissociation using the scan-along-the-track technique and obtained $\lambda$~=~$4.8^{+1.8}_{-1.1}$ m. The corresponding value in ordinary emulsion was found to be 10$^{+1.9}_{-1.4}$.\par
\indent By introducing lead salts into emulsion, we changed the relation between the numbers of Ar, Br and C, N, Î nuclei in such a way that the mean mass number did not increase for target nuclei of the last group ($\langle A\rangle~\cong$~47 and 44 for ordinary and enriched BR-2 emulsions, respectively). Using the data from Table 1, we can easily show that the observed considerable decrease (approximately by a factor of 2) in $\lambda$, for lead-enriched emulsion corresponds to the assumption that $^{12}$C dissociation on this nucleus (for which the cross section must increase in proportion to the target charge squared) is dominated by the Coulomb mechanism. This decrease is inconsistent with the hypothesis that purely diffractive mechanism governs this reaction in the entire mass-number region under consideration. The same situation takes place for coherent interaction of high-energy hadrons with nuclei (for example, for the reaction $\pi^{\pm}~\rightarrow~\pi^{\pm}\pi^+\pi^-)$.\par
\indent The values we obtained for the mean range with respect to coherent $^{12}$C dissociation in ordinary and lead-enriched emulsions correspond to cross sections of 20~$\pm$~4 and 43~$\pm$~16 mb/nucleus. The total cross sections for carbon dissociation into three fragments with $z$~=~2 (of course, these cross sections include contributions from incoherent mechanisms of this reaction $-$ for example, events of nuclear decay and so on) are significantly larger (several times according to \cite{Bondarenko}). Needless to say, we cannot specify all possible reaction channels with three fragments $^{3,4}$He in the final state and evaluate their cross sections on the basis of our data.\par
\indent Let us now consider the main features of relativistic $\alpha$ particles from the reaction $^{12}$C~$\rightarrow~3\alpha$ and compare them in the experiments with ordinary \cite{Abdurazakova} and lead-enriched emulsions. It should be emphasized that the initial energy of carbon nuclei and all experimental conditions (with the exception of target composition) were absolutely identical.\par
\indent Figure 1 shows integrated distributions in the square of the transverse momentum of $\alpha$ particles from the reaction $^{12}$C~$\rightarrow~3\alpha$ in both experiments. The transverse momentum $p_T$ was calculated by the formula $p_T~=~4p_0sin\theta$. This means that the analysis of $p_T$ distributions is essentially the analysis of the angular distributions of $\alpha$ particles. These angular distributions can be reliably measured by a photomethod owing to its high resolution.\par

\begin{figure}
    \includegraphics[width=4in]{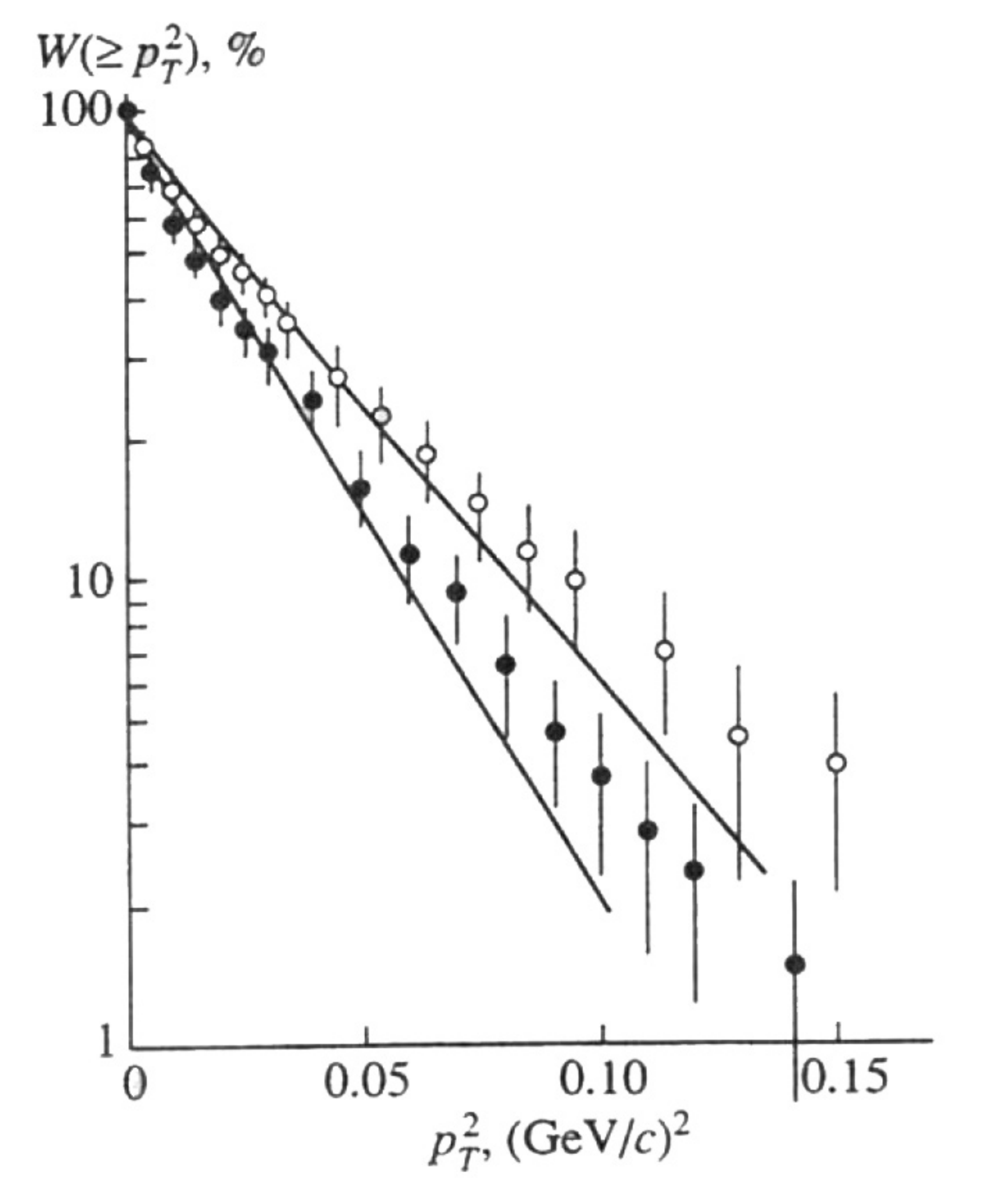}
    \caption{\label{Fig:1} $p^2_T$ distribution of $\alpha$ particles from the reaction $^{12}$C~$\rightarrow~3\alpha$ in ordinary ($\circ$,\cite{bvdklmt}) and lead-enriched ($\bullet$) emulsion. The straight lines represent distributions (1) at 2$\sigma^2=\langle p^2_T\rangle$.}
    \end{figure}

\indent Figure 1 shows that $p_T$ spectra of $\alpha$ particles are different in the two cases under consideration: the root-mean-square values $\langle p^2_T\rangle^{1/2}$ are equal to $192~\pm~10$ and $161~\pm~6$~MeV/$c$ for Em and Em~+~Pb, respectively (Table 2). It is also seen that both distributions shown in Fig. 1 are not consistent with the distribution\par
\indent $$d\sigma/dp^2_T~=~exp(-p^2_T/2\sigma^2)~~~~~(1)$$ \\$(2\sigma^2~=~\langle p^2_T\rangle)$ corresponding to partial normal distributions n(0, $\sigma$) in each of the transverse components of the $\alpha$-particle 3-momentum in the laboratory system [on the scale used in Fig. 1, distribution (1) is repre¬sented by a straight line]. \par
\indent However, any physical conclusions from these results would be premature, because the observed laboratory transverse momenta $p_T$ of the fragments of the projectile nucleus are distorted (increased) by its transverse motion (that is, by the transverse momentum $\mathbf q_T$ that is transferred to the nucleus undergoing fragmentation in the collision with a target \cite{Abdurazakova}). This circumstance is also manifested in our experiment $-$ in particular, in the distribution with respect to the pair azimuthal angle $\varepsilon_{ij}$~=~arccos($\mathbf p_{Ti}~\cdot~\mathbf p_{Tj}/p_{Ti}p_{Tj}$) between the transverse momenta $\mathbf p_{Ti}$ and $\mathbf p_{Tj}$ of a particles from one dissociation event. This distribution exhibits the azimuthal asymmetry of fragments: the distributions $d\sigma/d\varepsilon_{ij}$ peak at $\varepsilon_{ij}~\rightarrow~0$ with asymmetry coefficient\par
\indent $$A~=~(N_{\varepsilon_{ij}\ge\pi/2}-N_{\varepsilon_{ij}<\pi/2})/N_{0\le\varepsilon_{ij}\le\pi},~~~~~(2)$$ \\equal (Table 2) to $0.20~\pm~0.07$ (Em~+~Pb) and $0.21~\pm~0.09$ (Em), while the value of $A$ required by the momentum-conservation law in the decay into $N_{\alpha}$ particles, is equal to 1/($N_{\alpha}~-~1$)~=~0.5 and peaks at $\varepsilon_{ij}~\rightarrow~\pi$ (see Fig. 3 below). It is clear that the correct momentum values and correlation features can be obtained only upon going over to the rest frame of the nucleus undergoing dissociation.\par

\begin{table}
\caption{\label{}Comparative characteristics of relativistic $\alpha$ particles from the coherent reactions $^{12}$C~$\rightarrow~3\alpha$ in ordinary (Em) and lead-enriched (Em~+~Pb) emulsions at $p_0$~=~4.5~GeV/$c$ per nucleon.}
%\lable{Table:1}
\begin{tabular}{c|c|c}
\hline\noalign{\smallskip}
Characteristic & Em & Em~+~Pb  \\
\hline\noalign{\smallskip}
Number of $\alpha$ particles & 132 & 216 \\
$\langle p^2_T\rangle^{1/2}$, GeV/$c$ & $192~\pm~10$ & $161~\pm~6$ \\
$A$ (formula(2)) & $-0.21~\pm~0.09$ & $-0.20~\pm~0.07$ \\
$\langle p^{*2}_T\rangle^{1/2}$, GeV/$c$ & $141~\pm~7$ & $130~\pm~8$ \\
$A^*$ & $0.48~\pm~0.08$ & $0.43~\pm~0.06$ \\
$B^*$ (formula (5)) & $0.32~\pm~0.08$ & $0.44~\pm~0.06$ \\
$\langle p^2_{T,sum}\rangle^{1/2}$, GeV/$c$ & $383~\pm~42$ & $281~\pm~19$ \\
\hline\noalign{\smallskip}
\end{tabular}
\end{table}

\indent In the absence of additional (neutral) particles, this transformation is performed in a straightforward manner. As a result, we find that, at small angles of carbon scattering, the transverse momenta of $\alpha$ particles in the c.m.s. are given by\par
\indent $$\mathbf p^*_{Ti}~\cong~\mathbf p_{Ti}-\sum_{i=1}^3\mathbf p_{Ti}/3,~~~~~(3)$$ \\(here and below, asterisks label c.m.s. quantities). Figures 2 and 3 show the distributions in $p^{*2}_T$ and $\varepsilon^*_{ij}$~=~arccos($\mathbf p^*_{Ti}\cdot\mathbf p^*_{Tj}/p^*_{Ti}p^*_{Tj})$ for $\alpha$ particles from the reaction $^{12}$C~$\rightarrow~3\alpha$ in both experiments, and the Table 2 presents the main numerical characteristics of these distributions. These data lead to the following conclusions: \par
\indent (1)	As	might have been expected, the $\langle p^{*2}_T\rangle^{1/2}$ values are much less than $\langle p^2_T\rangle^{1/2}$ , but (in contrast to the latter) they coincide, for both experiments, within errors.\par
\indent (2)	Both distributions in Fig. 2 do not agree ($\chi^2$/NDF~=~2.8 and 4.2 for Em~+~Pb and Em, respectively) with distribution (1) written in the c.m.s.\par
\indent (3)	The distributions in $\varepsilon^*_{ij}$ are also inconsistent ($\chi^2$/NDF~=~1.9 and 1.7 for Em~+~Pb and Em, respectively) with the form \cite{Bondarenko2} \par
\indent $$d\sigma/d\varepsilon^*~\cong~\frac{1}{\pi}(1~+~c_1cos\varepsilon^*~+~c_2cos2\varepsilon^*),~~~~~(4)$$\\ which follows from the assumptions that each component of the 3-momentum of a particles in the c.m.s. obeys a normal distribution [normal partial distributions $n$(0, $\sigma$)] and that the energy-momentum is conserved in each decay event. The coefficients $c_1$ and $c_2$ in (4) are related to the azimuthal-asymmetry coefficients (2) (with the substitution of $\varepsilon_{ij}$ for $\varepsilon^*_{ij}$) and the azi-muthal-collinearity coefficients\par
\indent $$B^*~=~(N_{\varepsilon^*_{ij}\le\pi/4}+N_{\varepsilon^*_{ij}\ge3\pi/4}-N_{\pi/4<\varepsilon^*_{ij}<3\pi/4})/N_{0\le\varepsilon^*_{ij}\le\pi}~~~~~(5)$$ \\ by the equations\par
\indent $$c_1~=~-(\pi/2)A^*~=~-(\pi/2)(N_{\alpha}-1)^{-1},~c_2~=~-(\pi/2)B^*~=~(8\pi/25)(N_{\alpha}-1)^{-2}~~~~~(6)$$ \\ ($N\alpha$~=~3). Distribution (4) with coefficients (6) is shown in Fig. 3 (curve).\par

\begin{figure}
    \includegraphics[width=4in]{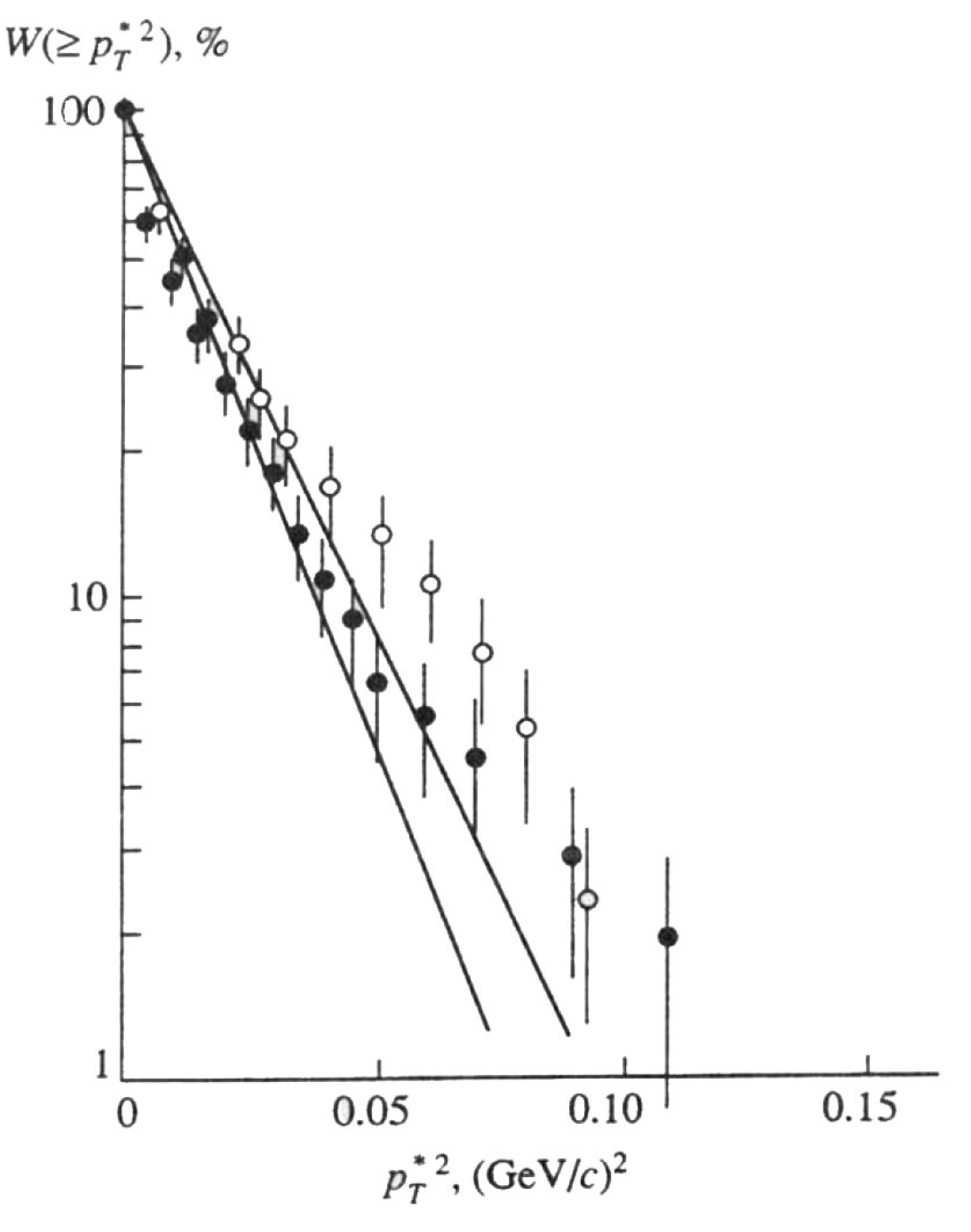}
    \caption{\label{Fig:2} As in Fig. 1, but for $p^{*2}_T$ distributions.}
\end{figure}

\begin{figure}
    \includegraphics[width=4in]{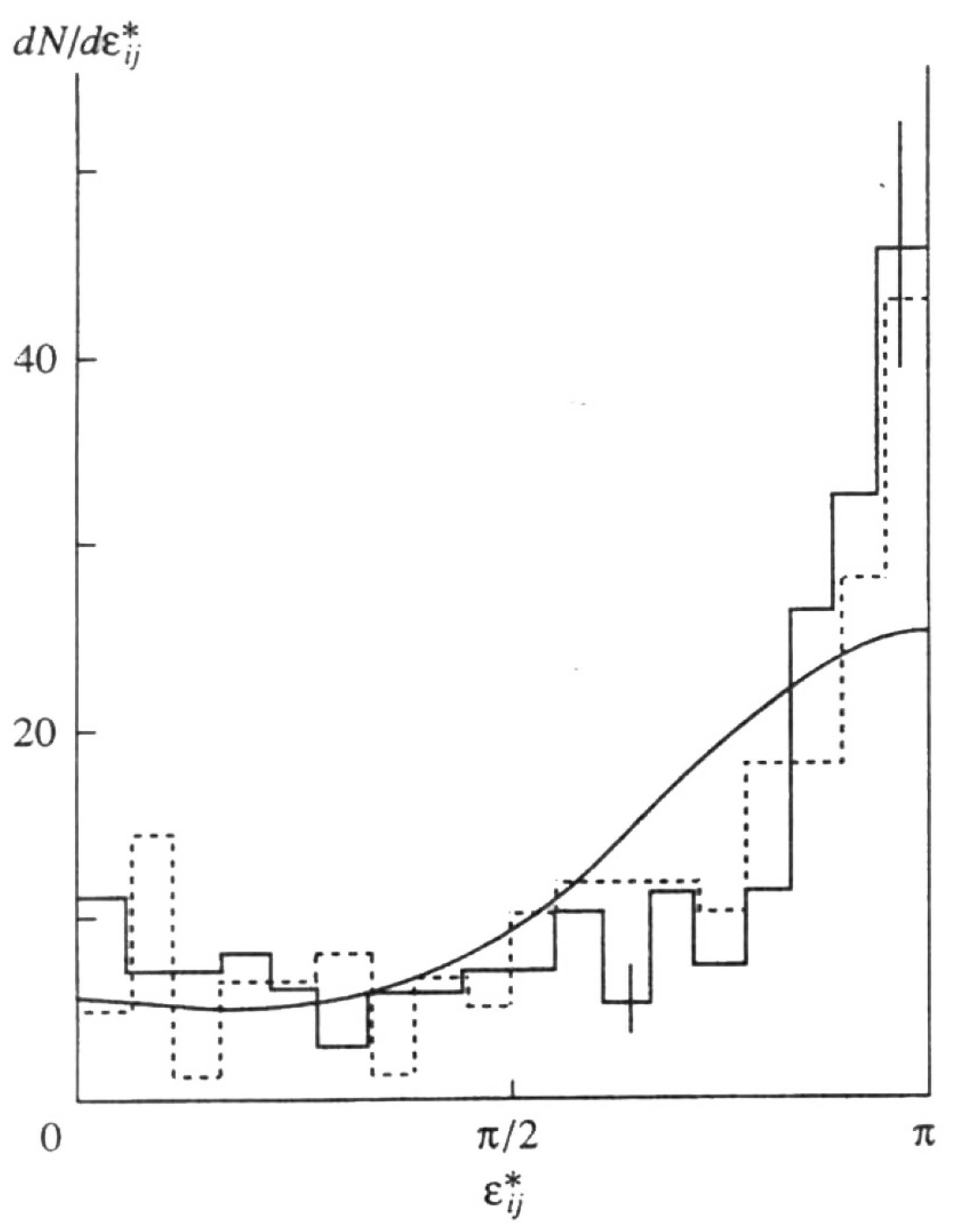}
    \caption{\label{Fig:3} $\varepsilon^*_{ij}$ distribution of $\alpha$ particles from the reaction $^{12}$C~$\rightarrow~3\alpha$: (curve) distribution (1) with coefficients (6) following from the phase-space model (normalization to the number of events in our experiment), (dashed histogram) Em, and (solid histogram) Em~+~Pb.}
\end{figure}
	
\indent The fact that the difference between the distributions in $p^2_T$ for the two sets of reactions is much more pronounced than the corresponding difference between the distributions in $p^{*2}_T$ means that the distributions in the transverse component $q_T$ of the momentum transfer to the carbon nucleus that undergoes dissociation must also differ. Indeed, the mean values $\langle q^2_T\rangle^1/2~(\mathbf q_T~=~\mathbf p_{T,sum}~=~\sum_{i=1}^{3}\mathbf p_{Ti}$) given in Table 2 differ significantly ($281~\pm~19$ and $383~\pm~42$~MeV/$c$ for Em~+~Pb and Em, respectively). This adds strength to the assumption that the dissociation $^{12}$C~$\rightarrow~3\alpha$ on Pb is dominated by the Coulomb mechanism. The $q^2_T$ distribution does not contradict (Fig. 4) distribution (1), which follows from the exponential dependence of $d\sigma/dt'$ on the 4-momentum transfer in a coherent collision of nuclei, because $|t$'$|~=~|t$~$-~t^{min}(M^*)|~\cong~q^2_T$, where $t^{min}(M^*)$ is the minimum 4-momentum realized at a minimum value of the effective mass $M^*~=\sum_{i}^{}m_i$ of projectile fragments.\par
\indent The narrowing of the $q^2_T$ distribution with increasing radius of the target nucleus occurs for the pure diffraction mechanism as well. This immediately follows from the coherence condition $qR~<~1$. However, the assumption that the contribution of the Coulomb mechanism to the reaction under study increases as we go over to the Em~+~Pb target seems more reasonable, because the mean mass number of the target in the lead-enriched stack does not exceed the mean mass number in the ordinary emulsion.\par
\indent In our experiment, the distributions $d\sigma/dM^*$ in the effective mass of three $\alpha$ particles from the reaction $^{12}$C~$\rightarrow~3\alpha$ can be obtained only under the assumption that their longitudinal momenta are equal to one another ($p_{Li}~\cong~p_0/3)$. In both cases (Em and Em~+~Pb), these distributions are smooth functions without any structures in the $M^*~-~3M\alpha$ range from zero to 40 MeV (Fig. 5). This supports the assumption that a large number of $^{12}$C states contribute to the formation of the final state of the reaction in question. Of course, richer experimental data and accurate momentum measure¬ments are required for the identification of these states.\par

\begin{figure}
    \includegraphics[width=4in]{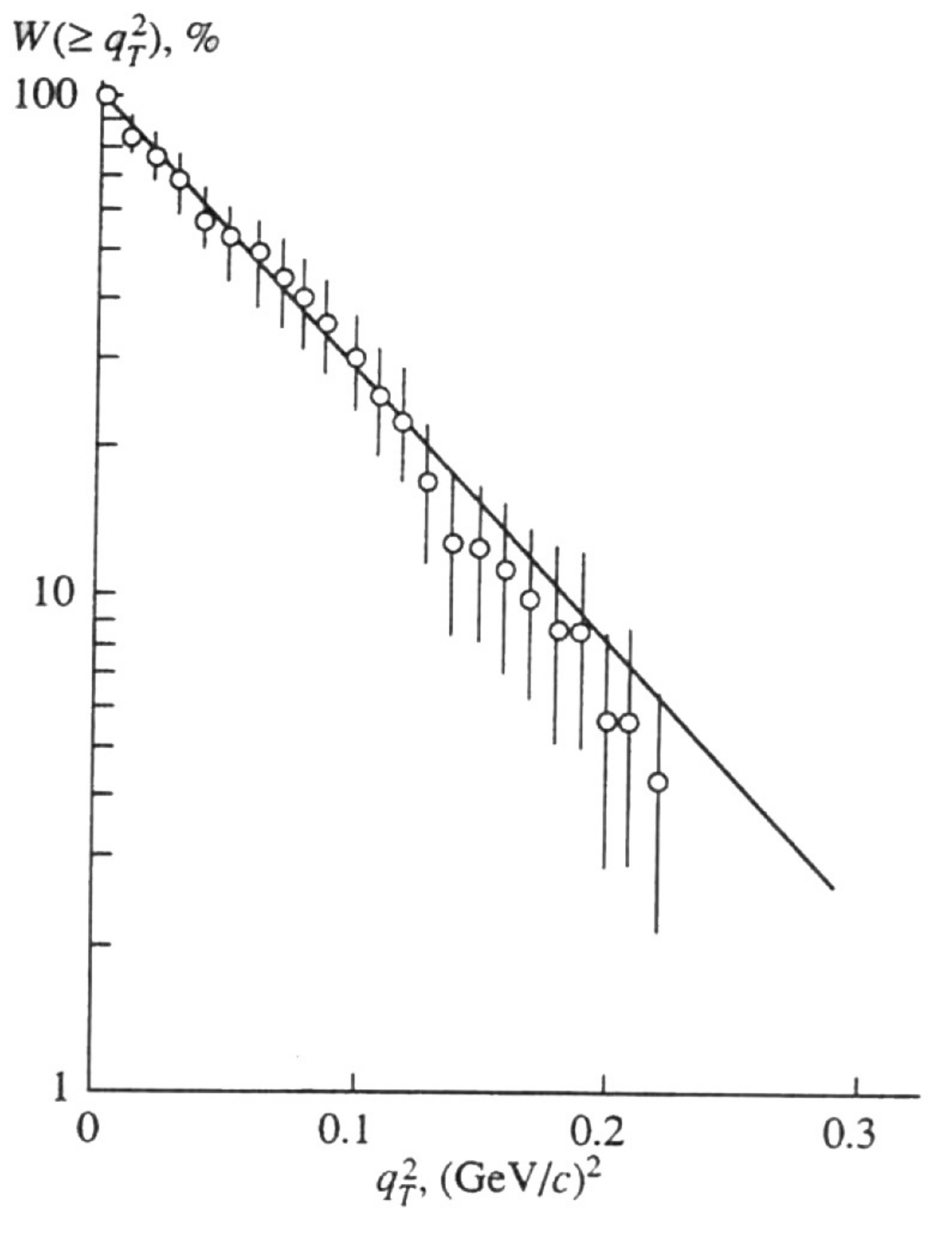}
    \caption{\label{Fig:4} $q^2_T~=~(\sum_{i=1}^{3}\mathbf p_{Ti})^2$ distribution for $^{12}$C~$\rightarrow~3\alpha$ events in our experiment.}
\end{figure}

\begin{figure}
    \includegraphics[width=4in]{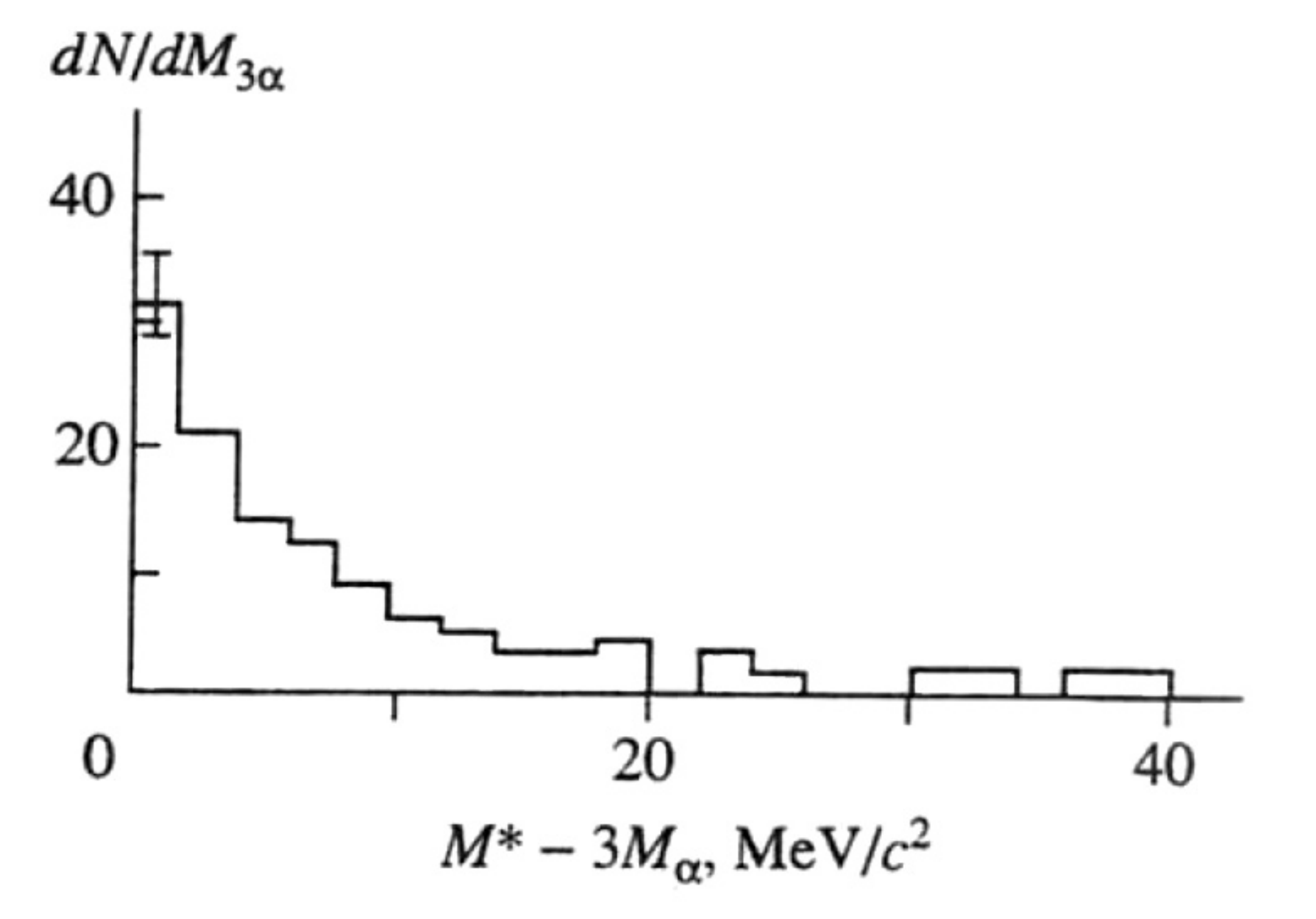}
    \caption{\label{Fig:5} Total distribution in the effective mass of three a particles from the reaction $^{12}$C~$\rightarrow~3\alpha$; the values of $M_{3\alpha}$ are shifted along the abscissa by the doubled $\alpha$ particle mass.}
\end{figure}

\indent However, the well-known statistical theory of fast fragmentation due to Feshbach, Huang, and Goldhaber \cite{Feshbach,Goldhaber} makes it possible to estimate the temperature of the carbon nucleus undergoing dissociation from the measured values of $\langle p^{*2}_T\rangle$. In $kT$ units, it is given by\par
\indent $$kT~=~\frac{A}{A-1}(\sigma^2_N/m_N),$$ \\ where\par
\indent $$\sigma^2_N~=~\sigma^2_{\alpha}(A-1)/A_{\alpha}(A-A_{\alpha})$$ \\(parabolic law), $A$~=~12, $A_{\alpha}$~=~4, $m_N$ is the nucleon mass, and $\sigma^2_{\alpha}~=~\langle p^{*2}_T\rangle/2$. Using the $\langle p^{*2}_T\rangle^{1/2}$ values from Table 2, we can easily obtain $kT$~=~3.4 and 4.0~MeV for lead-enriched and ordinary emulsions. The values obtained for $kT$ are significantly lower than those deter¬mined from the reactions of ordinary multifragmentation investigated in inclusive experiments $-$~that is, in reactions of the type $A~+~B~\rightarrow~\alpha~+~X$, where $A$ is a projectile nucleus, $B$ is a target nucleus, and $X$ stands for all other particles (see, for example \cite{Bengus,Bhanja,Bondarenko3,Adamovich}). This is true even with allowance for the fact that many authors overestimate $kT$, erroneously using the momentum characteristics of fragments in the laboratory system instead of the corresponding characteristics in the rest frame of the residual nucleus undergoing fragmentation. The values obtained here for $kT$ are also significantly lower than nucleon binding energy in a carbon nucleus. At the same time, the smallness of $kT$ is con¬sistent with a low energy-momentum transfer to the nucleus undergoing fragmentation, as is the case for coherent processes. The lowest $q^2_T$ value must correspond to Coulomb dissociation on Pb, which is probably observed.\par
\indent Deviations of $p^{*2}_T$ and $\varepsilon^*_{ij}$ distributions from expressions (1) and (4), which follow from the model of the fast statistical decay of the carbon nucleus indicate that this model, used to estimate its temperature, is unsatisfactory. The deviation of $d\sigma/dp^{*2}_T$ from the Rayleigh form (1) cannot be due to the complex composition of the target. Indeed, it can easily be shown that the total distribution must preserve its shape even if the parameter $\sigma^2$ in (1) explicitly depends on the target mass. However, such factors as the contribution of the cascade channel of carbon disintegration according to the scheme $^{12}$C~$\rightarrow~^8$Be~+~$\alpha~\rightarrow~3\alpha$, the possible nonzero angular momentum of the carbon nucleus undergoing dissociation, and the mechanisms of final-state interaction between a particles (identity effect) and between $\alpha$ particles and the target nucleus (rescattering) may be responsible for the deviations of $d\sigma/dp^{*2}_T$ and $d\sigma/d\varepsilon^*_{ij}$ from statistical expressions (1) and (4) [and, in particular, for the tendency toward a change in the slope of the $p^{*2}_T$ distribution in the region around 0.05~(GeV/$c)^2$ (Fig. 2)].\par
\indent To clarify the role of the cascade disintegration of carbon, we performed a Monte Carlo simulation of this process. The decays $^{12}$C~$\rightarrow~^8$Be~+~$\alpha$ and $^8$Be~$\rightarrow~2\alpha$ were assumed to occur in accordance with the statistical theory of fast fragmentation. The decay parameter was chosen in such a way that the mean momentum $\langle p^{*2}_T\rangle^{1/2}$ of a particles in the final state corresponded to the empirical value (Table 2). \par
\indent The calculated distributions $d\sigma/dp^{*2}_T$ and $d\sigma/d\varepsilon^*_{ij}$ for the direct and cascade versions of the decay $^{12}$C~$\rightarrow~3\alpha$ proved to be close to each other (they are not shown); hence, the available statistics of events are insufficient for determining the relative contributions of these channels. However, these statistics are quite sufficient for drawing the conclusion that the azimuthal-collinearity coefficient $B^*$ (see Table 2 and Fig. 3) exceeds the calculated values for both direct [$B^*$~=~0.16 (formula (6)] and cascade ($B^*$~=~0.18) versions. This excess is observed in both experiments. We believe that the only possible explanation of this result is that, in a collision that results in the dissociation of $^{12}$C into three $\alpha$ particles, not only energy and momentum, but also angular momentum is transferred to the projectile. Data reported in \cite{Babaev} indicate that angular momentum is transferred to the residual nucleus in ordinary multifragmentation as well.\par
\indent Let us summarize the main conclusions of this study.\par
\indent 1) Data on the decay $^{12}$C~$\rightarrow~3\alpha$ at $p_0$~=~4.5~GeV/$c$ per nucleon in ordinary and lead-enriched emulsions indicate that the coherent dissociation occurs. Diffraction mechanism is dominant for light and medium-mass target nuclei (see also \cite{Abdurazakova,Bondarenko}). In all probability, the Coulomb mechanism plays the principal role in the dissociation on the Pb nucleus because, in the lead-enriched emulsion, the mean cross section for $^{12}$C~$\rightarrow~3\alpha$ increases approximately by a factor of 2.\par
\indent 2) In the reaction under study, the decay temperatures of carbon nuclei depend only slightly on the mass number of the target nucleus. However, an increase in the contribution of the Coulomb mechanism of dissociation is accompanied by a significant decrease in the transverse-momentum transfer to the nucleus. This leads to the difference in the $p_T$ spectra of decay $\alpha$ particles in the laboratory system.\par
\indent 3) Temperatures obtained in this study ($kT~\approx~3.4~-~4.0$~MeV) are much lower than those in ordinary (non¬coherent) multifragmentation of relativistic projectile nuclei and are significantly lower than the nucleon binding energy in the nucleus undergoing disintegration.\par
\indent 4) The divergence of $\alpha$ particles produced in the reaction $^{12}$C~$\rightarrow~3\alpha$ in the transverse plane exhibits a tendency toward collinearity. In all probability, this is due to the angular-momentum transfer to the nucleus undergoing dissociation.\par

\begin{acknowledgments}
\indent The work of G.M.~Chernov was supported by the International Science Foundation.\par
\end{acknowledgments} 

%	\newpage

\end{document}